\documentstyle[12pt,graphicx,subfigure]{article}                
\begin{document}
\baselineskip 18pt
\def\today{\ifcase\month\or
 January\or February\or March\or April\or May\or June\or
 July\or August\or September\or October\or November\or December\fi
 \space\number\day, \number\year}
\def\thebibliography#1{\section*{References\markboth
 {References}{References}}\list
 {[\arabic{enumi}]}{\settowidth\labelwidth{[#1]}
 \leftmargin\labelwidth
 \advance\leftmargin\labelsep
 \usecounter{enumi}}
 \def\newblock{\hskip .11em plus .33em minus .07em}
 \sloppy
 \sfcode`\.=1000\relax}
\let\endthebibliography=\endlist
%
% A useful Journal macro
\def\Journal#1#2#3#4{{#1} {\bf #2}, #3 (#4)}

% Some useful journal names
\def\NCA{\em Nuovo Cimento}
\def\NIM{\em Nucl. Instrum. Methods}
\def\NIMA{{\em Nucl. Instrum. Methods} A}
\def\NPA{{\em Nucl. Phys.} A}
\def\NPB{{\em Nucl. Phys.} B}
\def\PLB{{\em Phys. Lett.}  B}
\def\PRL{\em Phys. Rev. Lett.}
\def\PRD{{\em Phys. Rev.} D}
\def\ZPC{{\em Z. Phys.} C}
\def\lsim{\ ^<\llap{$_\sim$}\ }
\def\gsim{\ ^>\llap{$_\sim$}\ }
\def\r2{\sqrt 2}
\def\beq{\begin{equation}}
\def\eeq{\end{equation}}
\def\beqn{\begin{eqnarray}}
\def\eeqn{\end{eqnarray}}
\def\rmuu{\gamma^{\mu}}
\def\rmud{\gamma_{\mu}}
\def\PL{{1-\gamma_5\over 2}}
\def\PR{{1+\gamma_5\over 2}}
\def\sinW2{\sin^2\theta_W}
\def\AEM{\alpha_{EM}}
\def\mul{M_{\tilde{u} L}^2}
\def\mur{M_{\tilde{u} R}^2}
\def\mdl{M_{\tilde{d} L}^2}
\def\mdr{M_{\tilde{d} R}^2}
\def\mz2{M_{z}^2}
\def\c2b{\cos 2\beta}
\def\au{A_u}
\def\ad{A_d}
\def\cob{\cot \beta}
\def\v#1{v_#1}
\def\tb{\tan\beta}
\def\epem{$e^+e^-$}
\def\KK{$K^0$-$\bar{K^0}$}
\def\wi{\omega_i}
\def\xj{\chi_j}
\def\Wmu{W_\mu}
\def\Wnu{W_\nu}
\def\m#1{{\tilde m}_#1}
\def\mH{m_H}
\def\mw#1{{\tilde m}_{\omega #1}}
\def\mx#1{{\tilde m}_{\chi^{0}_#1}}
\def\mc#1{{\tilde m}_{\chi^{+}_#1}}
\def\mwi{{\tilde m}_{\omega i}}
\def\mxi{{\tilde m}_{\chi^{0}_i}}
\def\mci{{\tilde m}_{\chi^{+}_i}}
\def\mz{M_z}
\def\sw{\sin\theta_W}
\def\cw{\cos\theta_W}
\def\cb{\cos\beta}
\def\sb{\sin\beta}
\def\rwi{r_{\omega i}}
\def\rxj{r_{\chi j}}
\def\rfp{r_f'}
\def\Kik{K_{ik}}
\def\Fq2{F_{2}(q^2)}
\def\mg{m_{\frac{1}{2}}}
\def\mchi1{m_{\chi}}
%%%%%%%%%%%%%%%%%%%%%%%%%%%%%%%%%%%%%%%%%%%%%%%%%%%%%%
%\def\sw5{\frac{\sin\theta_W}{\sqrt 2}}
%\def\cw5{-\frac{\cos\theta_W}{\sqrt 2}}
\def\tw{\tan\theta_W}
\def\sec2w{sec^2\theta_W}
%%%%%%%%%%%%%%%%%%%%%%%%%%%%%%%%%%%%%%%%%%%%%%%%%%%%%%

%%%%%%%%%%%%%%%%%%%%%%%%%%%%%%%%
\begin{center}{\Large \bf  
WMAP Constraints, SUSY Dark Matter and Implications for
the Direct Detection of SUSY}\\
\vskip.25in
{Utpal Chattopadhyay\footnote{E-mail: tpuc@iacs.res.in}$^{(a)}$, 
Achille Corsetti\footnote{E-mail: corsetti@neu.edu}$^{(b)}$ and  
Pran Nath\footnote{E-mail: nath@neu.edu}$^{(b)}$  }

{\it
(a) Department of Theoretical Physics, Indian Association for the Cultivation 
of Science, Jadavpur, Kolkata 700032, India\\
(b) Department of Physics, Northeastern University, Boston, MA 02115-5005, USA\\
}
\end{center}

\begin{abstract}  % DON'T CHANGE THIS LINE
Recently WMAP has measured the cosmological parameters to a much greater 
accuracy. We analyze the implications of this more precise measurement
for supersymmetric dark matter and for the direct detection of 
supersymmetry at accelerators. We consider mSUGRA including also
the hyperbolic branch (HB) in the radiative breaking of the electroweak
symmetry. On the part of the hyperbolic branch where the lightest neutralino 
is dominantly a higgsino rather than being mostly a bino, the relic density constraints
are satisfied by coannihilation with the next lightest neutralino and the light chargino. 
Including this branch the lightest neutralino mass satisfies 
$m_{\chi_1^0}\leq 1200$ GeV for $\tan\beta\leq 50$. 
Constraints of $b\rightarrow s+\gamma$, of $g_{\mu}-2$,
and of $B^0_{s}\rightarrow \mu^+\mu^-$ are also 
analyzed. It is shown that the neutralino-proton cross section 
in each case will fall within the reach of dark matter experiments.
Possibility for the direct detection of supersymmetry is discussed
in the allowed regions of the parameter space consistent with WMAP
constraints. A brief discussion of the hyperbolic branch and focus 
point region (HB/FP) is also given.
.\end{abstract}

\section{Introduction}
Recently the Wilkinson Microwave Anisotropy Probe (WMAP) has measured
some of the cosmological parameters with significantly greater 
precision\cite{bennett,spergel}. Specifically, WMAP gives the  
matter density of the universe so that $\Omega_m h^2 =0.135^{0.008}_{-0.009}$
and  gives the baryon density so that 
$\Omega_b h^2 =0.0224\pm 0.0009$,
where $\Omega_{m,b}=\rho_{m,b}/\rho_c$ where $\rho_{m,b}$ is the
matter (baryon) density  and $\rho_c$ is the mass density needed to 
close the universe and h is the Hubble parameter in units of 
100km/s/Mpc . 
Assuming the difference of the
two is cold dark matter (CDM) one finds the CDM density in the universe
according to WMAP is  now given by $\Omega_{CDM} h^2 =0.1126^{+0.008}_{-0.009}$.
In this paper we analyze the constraint of the WMAP results for
supersymmetric dark matter. For the analysis we will focus on the
mSUGRA model\cite{msugra} and analyze the allowed range of the
parameter space consistent with the WMAP relic density constraint.
 The above requires taking account of the full range of the hyperbolic 
branch of radiative breaking of the electroweak symmetry\cite{ccn}.
The mSUGRA model is characterized
by the parameters $m_0, m_{1/2}, A_0, \tan\beta$ where $m_0$ is the 
universal scalar mass, $m_{1/2}$ is the universal gaugino mass,
$A_0$ is the universal trilinear coupling and $\tan\beta$ is the
defined by $\tan\beta=<H_2>/<H_1>$ where $H_2$ gives mass to the up 
quark and the $H_1$ gives mass to the down quark and the lepton.
In the analysis  we will also consider the $b\rightarrow s\gamma$
constraint and the $g_{\mu}-2$ constraint. 
$\tan\beta$ in the 
analysis will range up to values of 50 and it is known\cite{ccn} that for
values of $\tan\beta$ which are large or even moderately large that
radiative breaking of the electroweak symmetry lies on the hyperbolic
branch. To make the discussion clearer  we review briefly
radiative breaking of the electroweak symmetry and discuss how the hyperbolic
branch arises in such a breaking. 
 One can illustrate this phenomenon analytically for the
case when the b quark couplings can be neglected. 
In this case one of the constraints of radiative symmetry breaking 
determines the Higgs mixing parameter $\mu$ so that\cite{ccn} 
\begin{equation}
	C_1m_0^2+C_3m'^2_{1/2}+C_2'A_0^2+\Delta \mu^2_{loop}=
	\mu^2+\frac{1}{2}M_Z^2
\end{equation}
Here, 
\begin{equation}
	m_{1/2}'=m_{1/2}+\frac{1}{2}A_0\frac{C_4}{C_3},~~
	C_2'=C_2-\frac{1}{4}\frac{C_4^2}{C_3}
\end{equation}
and, 
\beqn
C_1=\frac{1}{t^2-1}(1-\frac{3 D_0-1}{2}t^2), C_2=\frac{t^2}{t^2-1}k\nonumber\\
~C_3=\frac{1}{t^2-1}(g- t^2 e), 
~C_4=-\frac{t^2}{t^2-1}f,
\Delta \mu^2_{loop}=\frac{\Sigma_1- t^2\Sigma_2}{t^2-1}
\eeqn
$\Delta \mu^2$ is the loop correction. $\Sigma_{1,2}$ is  
as defined in Ref.\cite{ccn} , $t=\tan\beta$ and the functions $e,f,g,k$ are 
as defined in Ref.\cite{ilm}. Further,  $D_0=1-(m_t/m_f)^2$ and 
$m_f\simeq 200 sin\beta$~GeV. 

	For small to moderate values of $\tan\beta$ 
 the loop corrections
	are typically small and further the renormalization group 
	analysis shows that $C_2'>0$ and $C_3>0$. For such values of 
$\tan\beta$ where the loop corrections have reduced scale dependence 
one finds $C_1>0$ independent of any scale choice $Q$ for having the 
radiative electroweak symmetry breaking (EWSB).   
        In this circumstance one finds that the radiative
	symmetry breaking constraint demands that the allowed set
	of soft parameters $m_0$ and 
$m_{\frac{1}{2}}'$ for a given value of $\mu$ lie 
	on the surface of an ellipsoid.  
  This condition then 
	places an upper bound on sparticle masses for a given value of 
        $\Phi$ which is the fine tuning parameter defined by  
	$\Phi =\frac{\mu^2}{M_Z^2} +\frac{1}{4}$\cite{ccn}.  
	This is the ellipsoidal branch of radiative breaking of the
	electroweak symmetry\cite{ccn}. 
	However, it was found in Ref.\cite{ccn} that for typically 
larger $\tan\beta$ $(\gsim 7)$ 
	when the loop corrections to $\mu$ are significant along with 
        a significant degree of its variation with the scale $Q$, 
        the above 
	scenario does not necessarily hold.  One way to see this
	phenomenon is to choose a value of the running scale 
$Q_0$ at which the 
	loop corrections to $\mu$ are minimized. One finds then
	that in some parts of the parameter space where $m_0$ and 
$m_{1/2}$ are relatively larger the minimization scale $Q_0$ occurs 
in such a region that it leads to
	a switch in the sign of $C_1$, i.e. sign($C_1(Q_0)$)=$-1$.
	In this circumstance one finds that the radiative 
	symmetry breaking condition takes the form 	
	\begin{equation}
\frac{m_{1/2}'^2}{\alpha^2(Q_0)}-\frac{m_0^2}{\beta^2(Q_0)}\simeq \pm 1
\end{equation}
where the sign $\pm$ is determined by the condition 
$sign ((\Phi+\frac{1}{4})M_Z^2-C_2'A_0^2)=\pm$  and where 
\beq
\alpha^2=\frac{|(\Phi_0+\frac{1}{4})M_Z^2-C_2'A_0^2|}{|C_3|},~~~
\beta^2=\frac{|(\Phi_0+\frac{1}{4})M_Z^2-C_2'A_0^2|}{|C_1|}
\eeq
From the above we see that the presence of the relative minus sign leads to
a drastically different constraint on the soft parameters due to constraint
of the radiative breaking of the electroweak symmetry. Here for
fixed  values of $A_0$ one finds that $m_0$ and $m_{\frac{1}{2}}'$ lie 
on a hyperbola and thus these 
parameters can get large for fixed values of $\mu$ or for fixed values
of the fine tuning parameter $\Phi$. This is the high zone of the
hyperbolic branch
of radiative breaking of the electroweak symmetry\cite{ccn}. Remarkably, 
the soft parameters can be quite large even while the value of 
$\Phi$ or $\mu$ can be chosen to be significantly small.  
This is a feature which gives a significantly different type   
of mixing of gauginos and higgsinos than the usually explored regions 
of the minimal supergravity model. In the high zone of the hyperbolic
branch when   $m_{\frac{1}{2}}>>\mu$, an inversion phenomenon takes 
 place, and the neutralino mass becomes essentially $\mu$. 
The above has a drastic effect on sparticle spectrum and on supersymmetry
 phenomenology which we discuss below.

\section{Sparticle Spectrum in the Inversion Region of the Hyperbolic Branch}
As discussed in Sec.1 the constraints on $m_0$ and $m_{\frac{1}{2}}$
for fixed $\mu$ for the hyperbolic branch 
are very different than for the usual (ellipsoidal)
scenario. Here since $m_0$ and $m_{\frac{1}{2}}$ 
can get large for fixed $\mu$ one finds that the  squark and slepton
masses get very heavy and may lie in the several TeV range 
(The feature of large $m_0$ is shared by the focus point region of 
mSUGRA models\cite{fmm}).  We consider here a specific part of the
hyperbolic branch where $m_{\frac{1}{2}}>>\mu >> M_Z$.
In this scenario then one finds
the two lightest neutralino states $\chi_1^0, \chi_2^0$ 
and the light chargino state $\chi_1^{\pm}$ are essentially degenerate, 
each with mass $\sim |\mu|$. We will call this 
phenomenon "inversion" in that the lightest neutralino switches from
being mostly a Bino to being purely a higgsino. In fact, this is also 
the case for the second lowest neutralino and the lighter chargino
since all of them have a common mass $\mu$ to the leading order.
The degeneracy in lifted when corrections $O(M_Z^2/M_{1,2})$
and $O(M_Z^2/\mu)$ are included.
 The remaining sparticle spectrum consisting of quarks,
sleptons, gluino and the remaining charginos and neutralinos are
significantly higher and in principle could lie in the several TeV
range and perhaps beyond the reach of even the LHC.
Thus the prospects of observing supersymmetry depends on our
ability to observe the particles $\chi_1^0, \chi_2^0$ and 
$\chi_1^{\pm}$ in addition to the observation of the light
Higgs boson. 
Including the lowest order perturbation corrections
$O(M_Z^2/M_{1,2})$ and $O(M_Z^2/\mu)$ 
the masses of these three lowest mass states in the inversion
region at the tree level are given by
\beqn
M_{\chi_1^0}= \mu -\frac{M_Z^2}{2} (1-\sin 2\beta) 
[\frac{\sin^2\theta_W}{M_1-\mu} + \frac{\cos^2\theta_W}{M_2-\mu}]\nonumber\\
M_{\chi_2^0}= \mu +\frac{M_Z^2}{2} (1+\sin 2\beta) 
[\frac{\sin^2\theta_W}{M_1+\mu} + \frac{\cos^2\theta_W}{M_2+\mu}]\nonumber\\
M_{\chi_1^{\pm}}= \mu +\frac{M_W^2 \cos^2 \beta}{\mu}
-\frac{M_W^2}{\mu} \frac{(M_2\cos\beta + \mu \sin\beta)^2}{(M_2^2-\mu^2)}
\eeqn
Thus for $\mu>0$ the mass pattern that emerges is 
\beq
m_{\chi_1^0}< m_{\chi_1^{\pm}}<m_{\chi_2^{0}}
\eeq
The quantities that are relevant for the observability
of these sparticles are the mass differences 
\beq
\Delta M^{\pm} = m_{\chi_1^{\pm}}- m_{\chi_1^0},
~~
\Delta M^{0} = m_{\chi_2^{0}}- m_{\chi_1^0}
\eeq
While $m_{\chi_1^{\pm}}$,  $m_{\chi_1^{0}}$,   and $m_{\chi_1^0}$
masses lie in the several hundred GeV to TeV (above TeV) 
range the mass differences
$\Delta M^{\pm}$ are much smaller and lie in the range 1-10 GeV.  
The mass differences can receive loop corrections\cite{pierce,drees} which can be 
as much as 25\% or more. However, these corrections do not modify 
the general picture of this scenario.
The above leads to some important constraints on what  
may be observed experimentally. 
\section{Coannihilation, relic density, and detection rates 
with WMAP Constraints}
We discuss now the WMAP constraints on SUSY dark matter and also 
investigate if such dark matter will be accessible to direct
detection.  This issue is of great importance as there are
 on going dark matter experiments\cite{dama,cdms,hdms,edelweiss}
and also experiments planned for the future\cite{genius,cline}
to detect dark matter. In the analysis we will use a $2\sigma$
constraint on the WMAP\cite{bennett,spergel} result for CDM, i.e., we take
\beq
\Omega_{\chi}h^2 = 0.1126^{+0.016}_{-0.018}
\label{wmapeqn}
\eeq
Many interesting theoretical investigations in the analysis of 
supersymmetric dark matter have been carried out over the 
years\cite{direct,drees,baer,chatto}.
These include investigations of the
effects of the variations of uncertainties in the relic density and
wimp velocity on the detection rates\cite{bottino1}, effects of 
 nonuniversalities in the Higgs sector\cite{nonuni,nonuni1}  
 and in the gaugino sector\cite{corsetti2,nelson}, effects of 
 CP phases\cite{cin}, and the effects of Yukawa 
 unification\cite{gomez,corsetti3}.
  More recently the effects of coannihilation on
 supersymmetric dark matter have been 
analyzed\cite{mizuta,efo1,efo2,gomez,ads,efo3,nihei,bednyk}.
This effect becomes important
when the mass of the next to the lightest supersymmetric particle  
(nlsp) is close 
to the mass of the lightest supersymmetric particle (lsp)
at the time when the lsp's decouple from the background. In such 
a situation the coannihilation processes involving lsp-nlsp and the nlsp-nlsp 
annihilation 
must be taken into account.
The quantity of interest is the number density $n_a=\sum n_a$ where $a$ 
runs over
the particle types that enter in coannihilation, and $n$ obeys the 
Boltzmann equation
\beqn
\frac{dn}{dt}= -3H n -<\sigma_{eff} v> (n^2-n_0^2)
\eeqn
where H is the Hubble parameter, $n_0$ is the equilibrium number 
density and $\sigma_{eff}$ is the effective total cross section 
defined by 
\beq
\sigma_{eff}= \sum\sigma_{ab} r_a r_b
\eeq
where $\sigma_{ab}$ is the annihilation cross section of particle a with
particle b, and $r_a=n_{0a}/n_0$ where $n_{0a}$ is the density  of
 particles of species $a$ at equilibrium. After the freeze out the 
 nlsp's decay to the lsp and thus n becomes the number density of the lsp. 
It was shown that in mSUGRA one naturally has coannihilation 
with the sleptons when the neutralino mass extends to masses 
beyond 150-200 GeV with processes of the type
$\chi \tilde \ell_R^a \rightarrow \ell^a \gamma, \ell^a Z, \ell^a h$,
$\tilde \ell_R^a \tilde \ell_R^b \rightarrow \ell^a \ell^b$,
 and $\tilde \ell_R^a \tilde \ell_R^{b*} \rightarrow \ell^a\bar \ell^b,
\gamma \gamma, \gamma Z, ZZ, W^+W^-, hh$ where $\tilde{\it l}$ is 
essentially a $\tilde \tau$. The above coannihilation
processes extend the allowed neutralino range up to 700 GeV\cite{efo2}.
We will show that remarkably the relic density constraints can
be satisfied on the hyperbolic branch also by coannihilation. 
However, on the hyperbolic branch the  coannihilation is of
 an entirely different nature. Specifically in the inversion region
 the dominant coannihilation is the  
$ \chi_1^0- \chi_1^{\pm}$ coannihilation followed by 
$\chi_1^0-{\chi_2^0}$ coannihilation, and by
$\chi_1^{+}- \chi_1^{-}$  and by $\chi_1^{\pm}-\chi_2^0$ 
coannihilations. Some of the dominant processes that contribute
to the above coannihilation processes are\cite{Edsjo:1997bg}
\beqn
\chi_1^0 \chi_1^{+}, \chi_2^0 \chi_1^{+}\rightarrow u_i\bar d_i, \bar e_i\nu_i, AW^+,Z W^+,
W^+h \nonumber\\
\chi_1^{+} \chi_1^{-}, \chi_1^0 \chi_2^{0}\rightarrow u_i\bar u_i, d_i \bar d_i, 
W^+W^-
\label{charginocoani}
\eeqn 
Since the mass difference between the states 
$\chi_1^+$ and $\chi_1^{0}$ is the smallest the $\chi_1^0 \chi_1^{+}$
coannihilation dominates. 

\begin{figure}                       
\vspace*{-1.0in}                                 
\subfigure[A plot in the $m_0-m_{\frac{1}{2}}$ plane of the allowed region 
 consistent with electroweak symmetry breaking and the WMAP relic density 
 constraints for the mSUGRA case.
The input parameters are $A_0=0, \tan\beta =10, \mu>0$ and the
relic density constraint imposed is of Eq.(\ref{wmapeqn}).
The white region is the parameter space allowed by the electroweak
symmetry breaking constraints while the shaded region is disallowed.
The filled circles denote the region allowed by the relic density constraint.
The filled circles just below the upper shaded region 
arise from  the hyperbolic branch. 
$a_\mu^{SUSY}(-1.5\sigma)$ contour is the 
black line.]{
\label{t10_m0_mhalf_msugra}             
\hspace*{-0.4in}                               
\begin{minipage}[b]{\textwidth}                       
\centering                      
\includegraphics[width=0.7\textwidth]{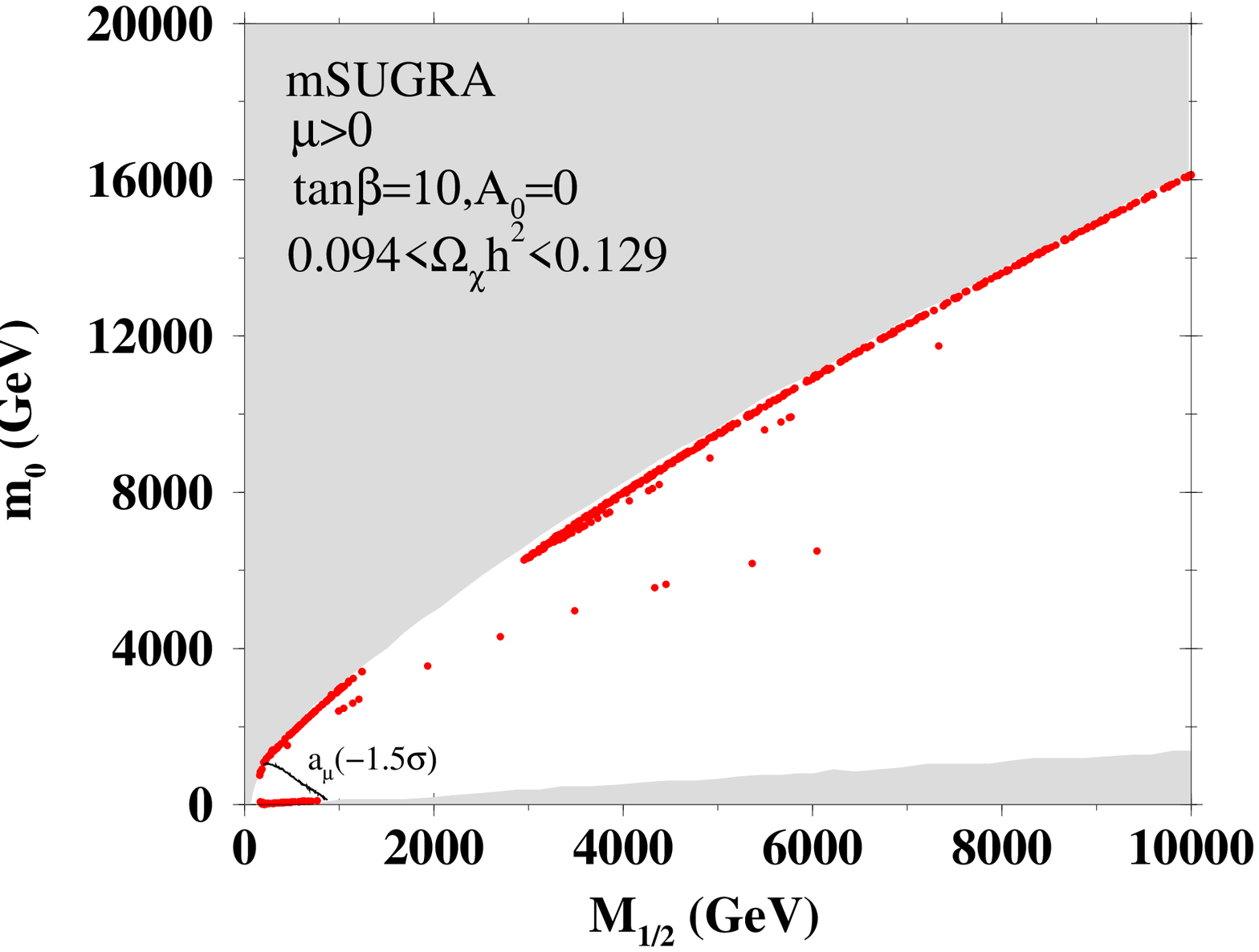} 
\end{minipage}}                       

%%%%%%%%%%%%%%%%%%%%%%%%%%%%%%%%%%%%%%%%%%%%%%%%%%
%%%%%%%%%%%%%%%
\subfigure[A plot in the $m_0-m_{\chi_1}$ plane of the allowed region 
represented by black circles  consistent with electroweak symmetry
 breaking and   WMAP relic density 
 constraints of Eq.(\ref{wmapeqn}) for the mSUGRA case including the parameter 
 space on the hyperbolic branch. The input parameters are the same as in 
 Fig.~\ref{t10_m0_mhalf_msugra}]{
\label{t10_m0_mchi_msugra}                      
\hspace*{-0.6in}                     
\begin{minipage}[b]{0.5\textwidth}                       
\centering
\includegraphics[width=\textwidth]{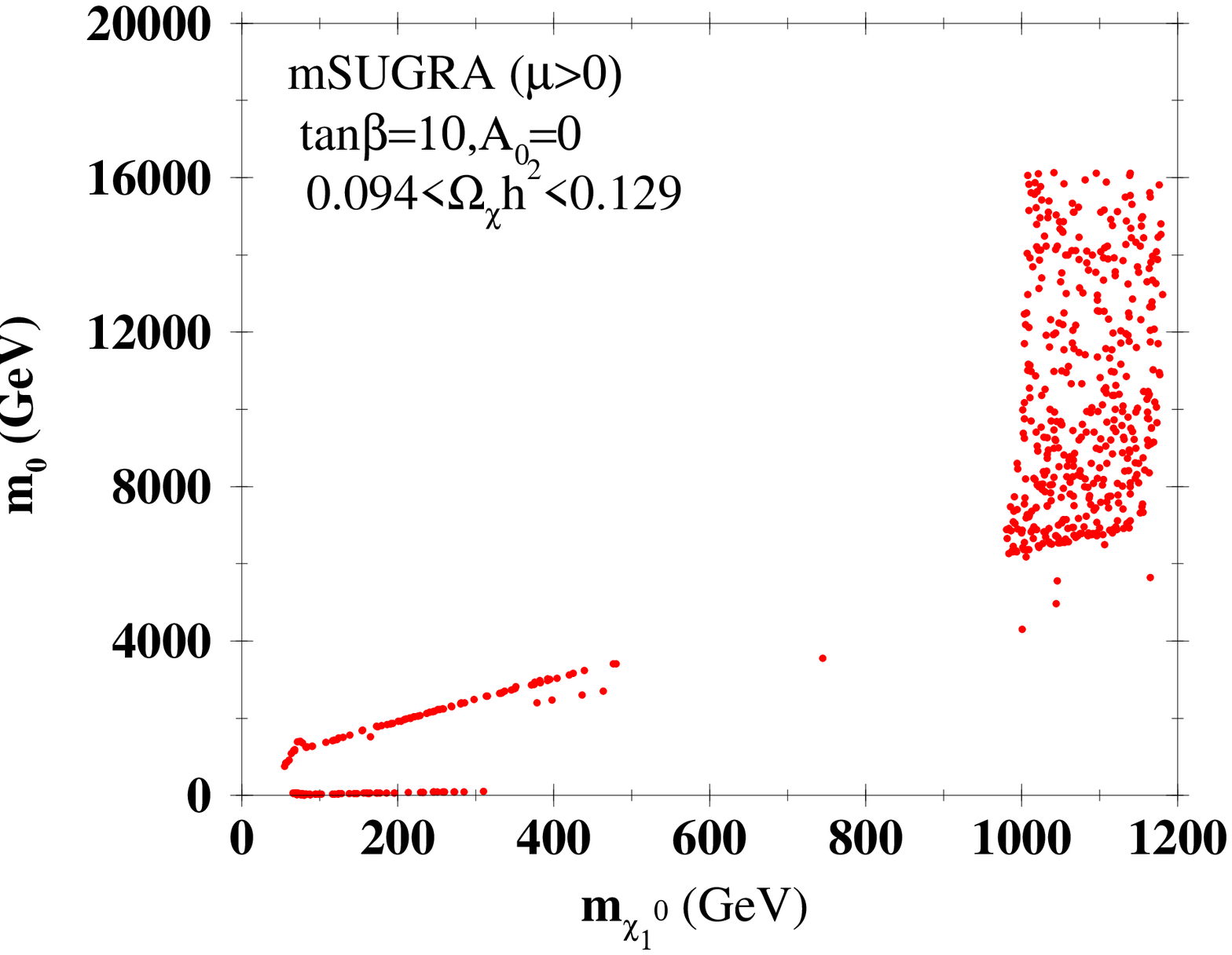}
\end{minipage}}
\hspace*{0.3in}                       
\subfigure[Same as Fig.~\ref{t10_m0_mchi_msugra} except that the
plot is in the $m_{\frac{1}{2}}-m_{\chi_1^0}$ plane.]{
\label{t10_mhalf_mchi_msugra}                       
\begin{minipage}[b]{0.5\textwidth}                       
\centering                      
\includegraphics[width=\textwidth]{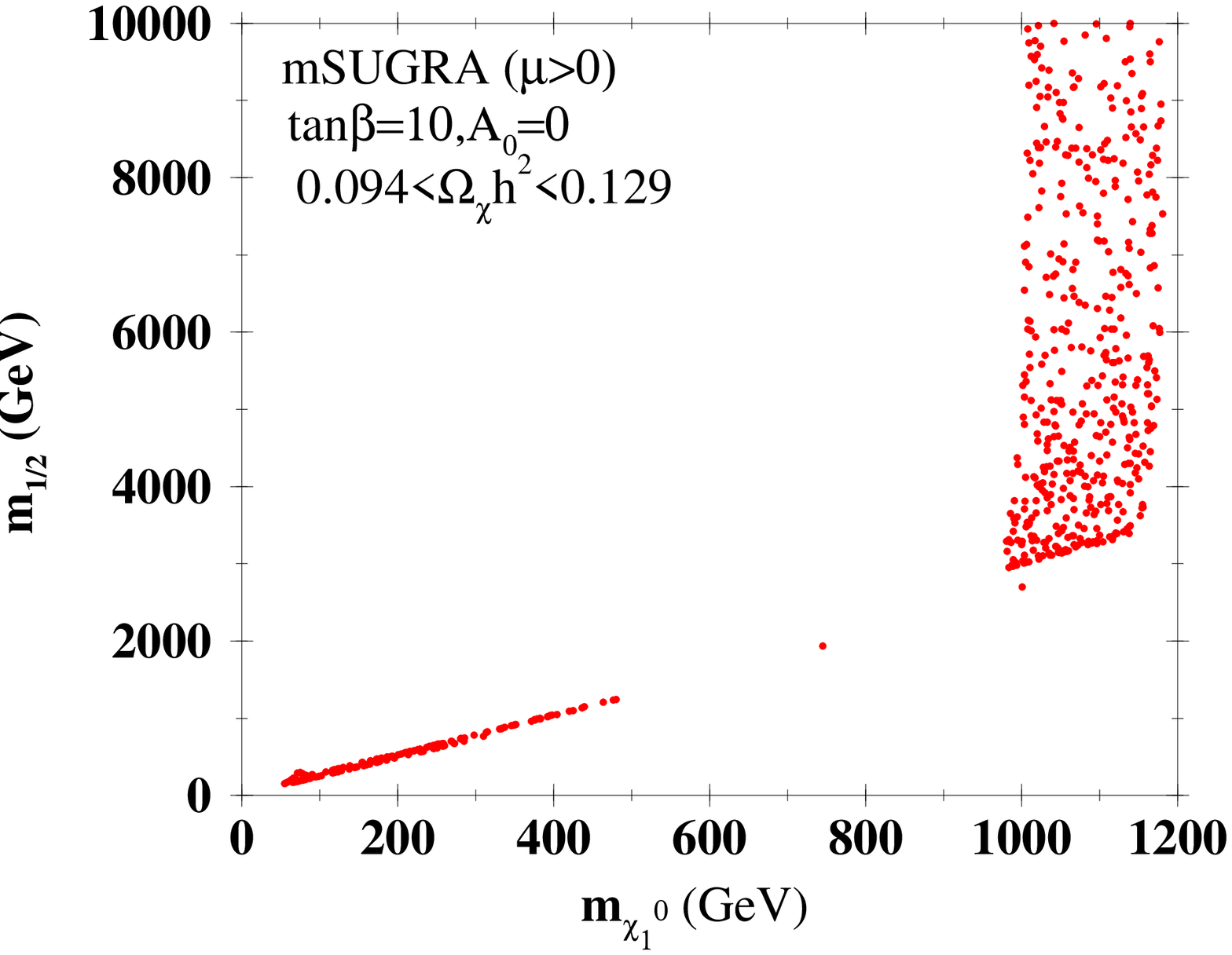}
\end{minipage}}                       
\caption{Relic density constraint and neutralino mass range for $\tan\beta=10$}
\label{tan10threefigs}
\end{figure}

\begin{figure}                       
\vspace*{-1.0in}                                 
\subfigure[Same as Fig.~\ref{t10_m0_mhalf_msugra} 
except $\tan\beta =30$. $b \rightarrow s+ \gamma$ contour and 
$a_\mu^{SUSY}$ contours are also shown.]{
\label{t30_m0_mhalf_msugra}             
\hspace*{-0.4in}                               
\begin{minipage}[b]{\textwidth}                       
\centering                      
\includegraphics[width=0.7\textwidth]{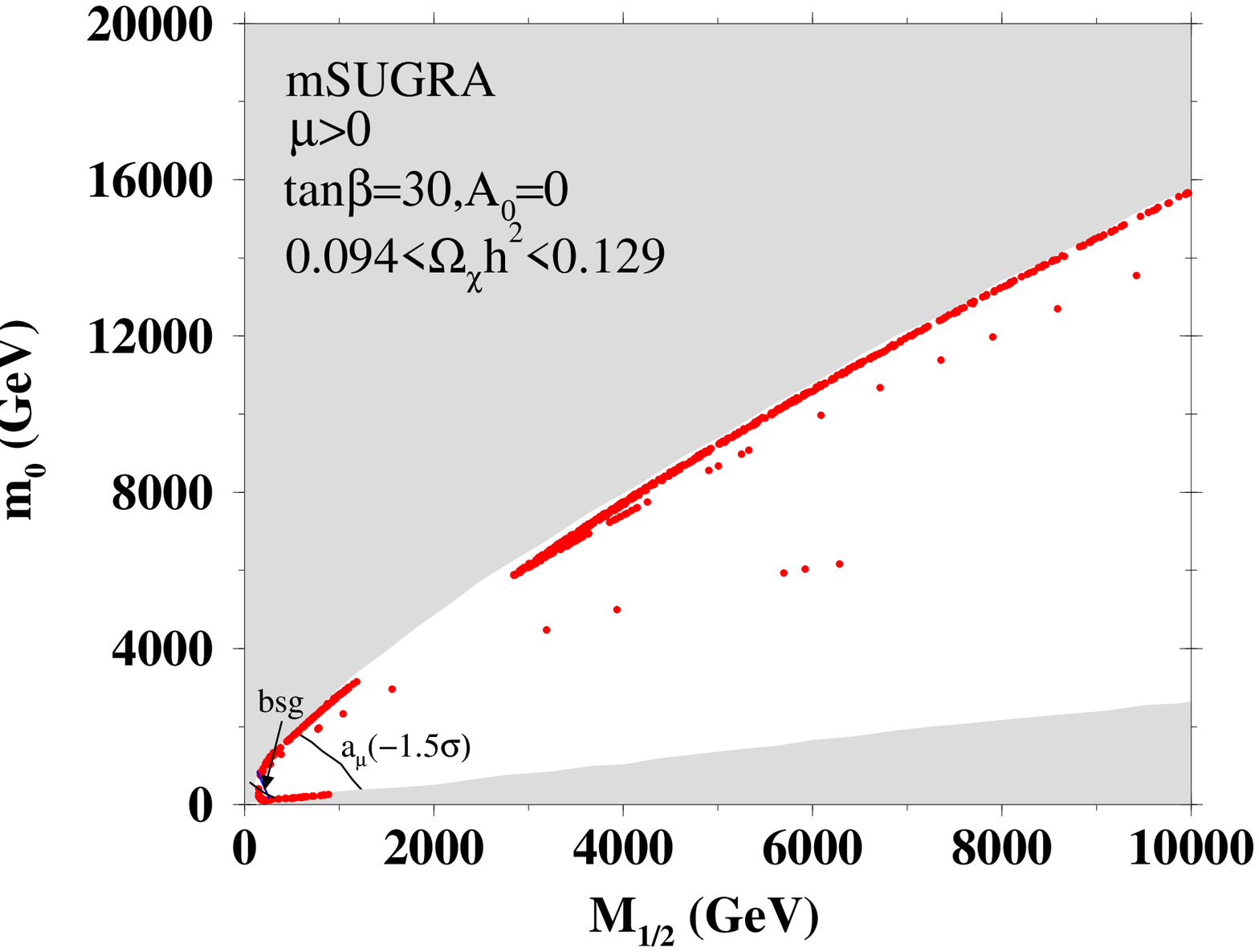} 
\end{minipage}}                       

%KEEP A LINE BLANCK AS ABOVE IN ORDER TO HAVE THE FOLLOWING TWO FIGS AT 
% THE BOTTOM 
\subfigure[Same as Fig.~\ref{t10_m0_mchi_msugra} 
except  $\tan\beta =30$.]{
\label{t30_m0_mchi_msugra}                      
\hspace*{-0.6in}                     
\begin{minipage}[b]{0.5\textwidth}                       
\centering
\includegraphics[width=\textwidth]{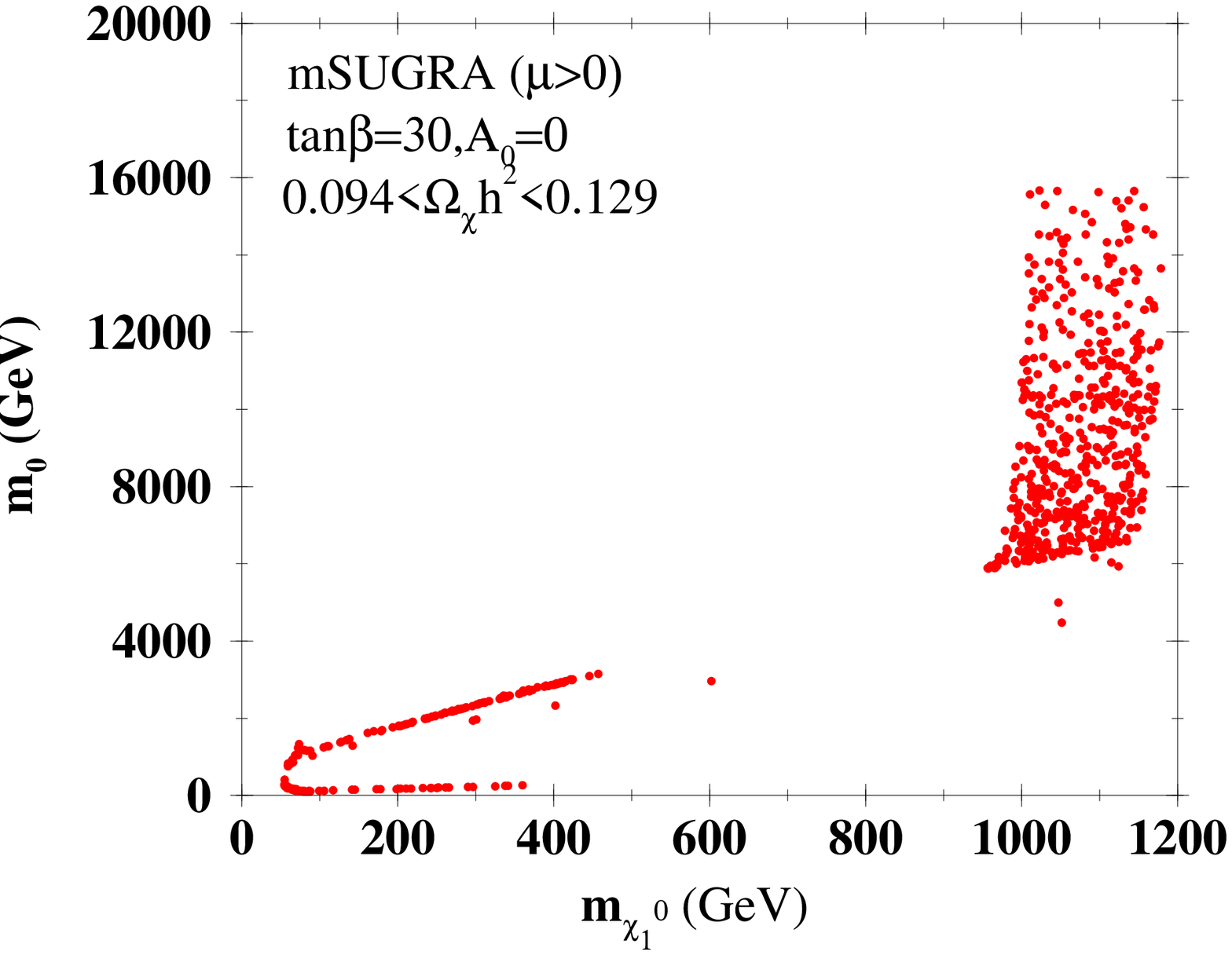}
\end{minipage}}
\hspace*{0.3in}                       
\subfigure[Same as Fig.~\ref{t10_mhalf_mchi_msugra} 
except $\tan\beta =30$.]{
\label{t30_mhalf_mchi_msugra}                       
\begin{minipage}[b]{0.5\textwidth}                       
\centering                      
\includegraphics[width=\textwidth]{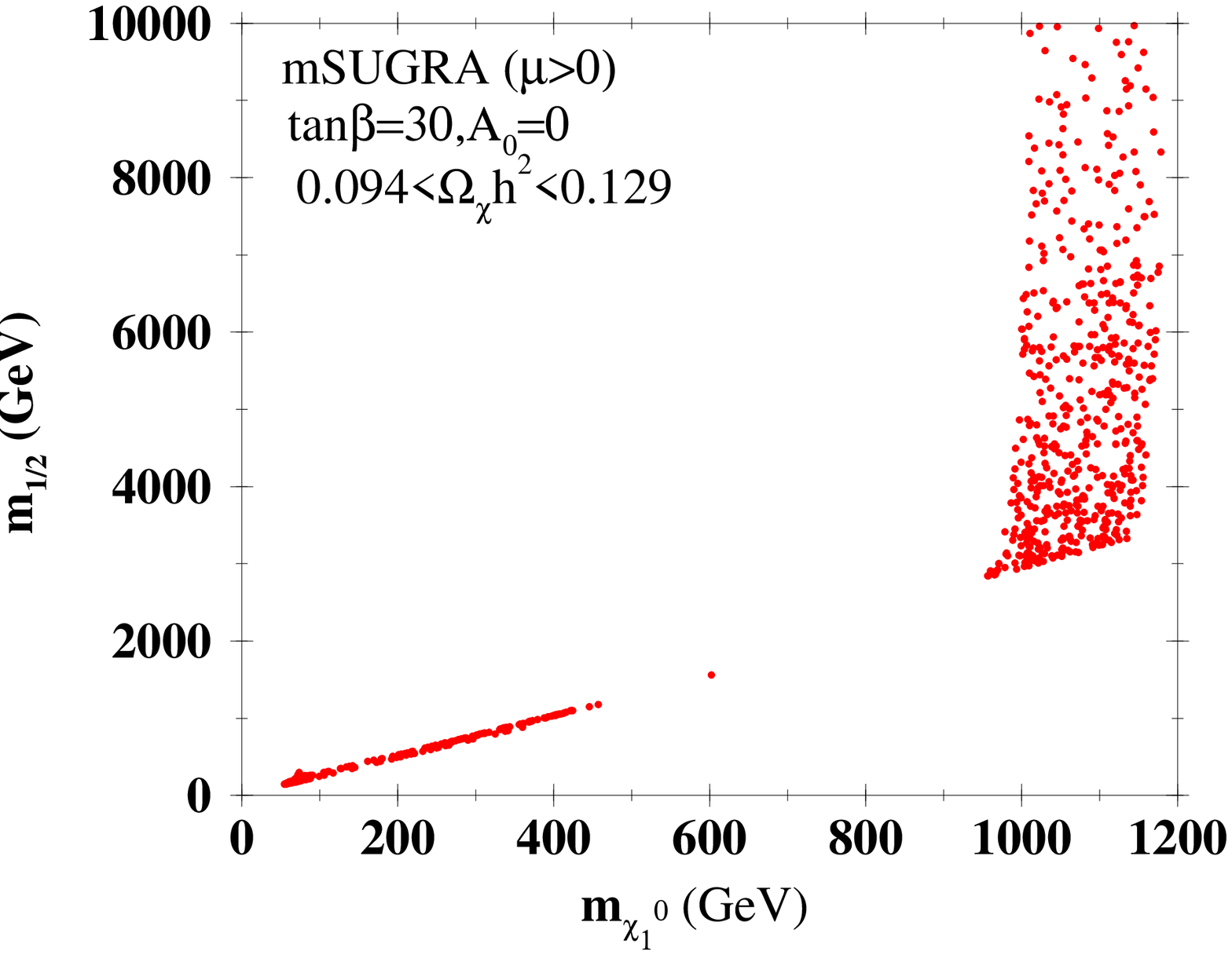}
\end{minipage}}                       
\caption{Relic density constraint and neutralino mass range for $\tan\beta=30$}
\label{tan30threefigs}
\end{figure}

\begin{figure}                       
\vspace*{-1.0in}                                 
\subfigure[Same as Fig.~\ref{t10_m0_mhalf_msugra} 
except $\tan\beta =50$. $b \rightarrow s+ \gamma$ contour and 
$a_\mu^{SUSY}$ contours are also shown.]{
\label{t50_m0_mhalf_msugra}             
\hspace*{-0.4in}                               
\begin{minipage}[b]{\textwidth}                       
\centering                      
\includegraphics[width=0.7\textwidth]{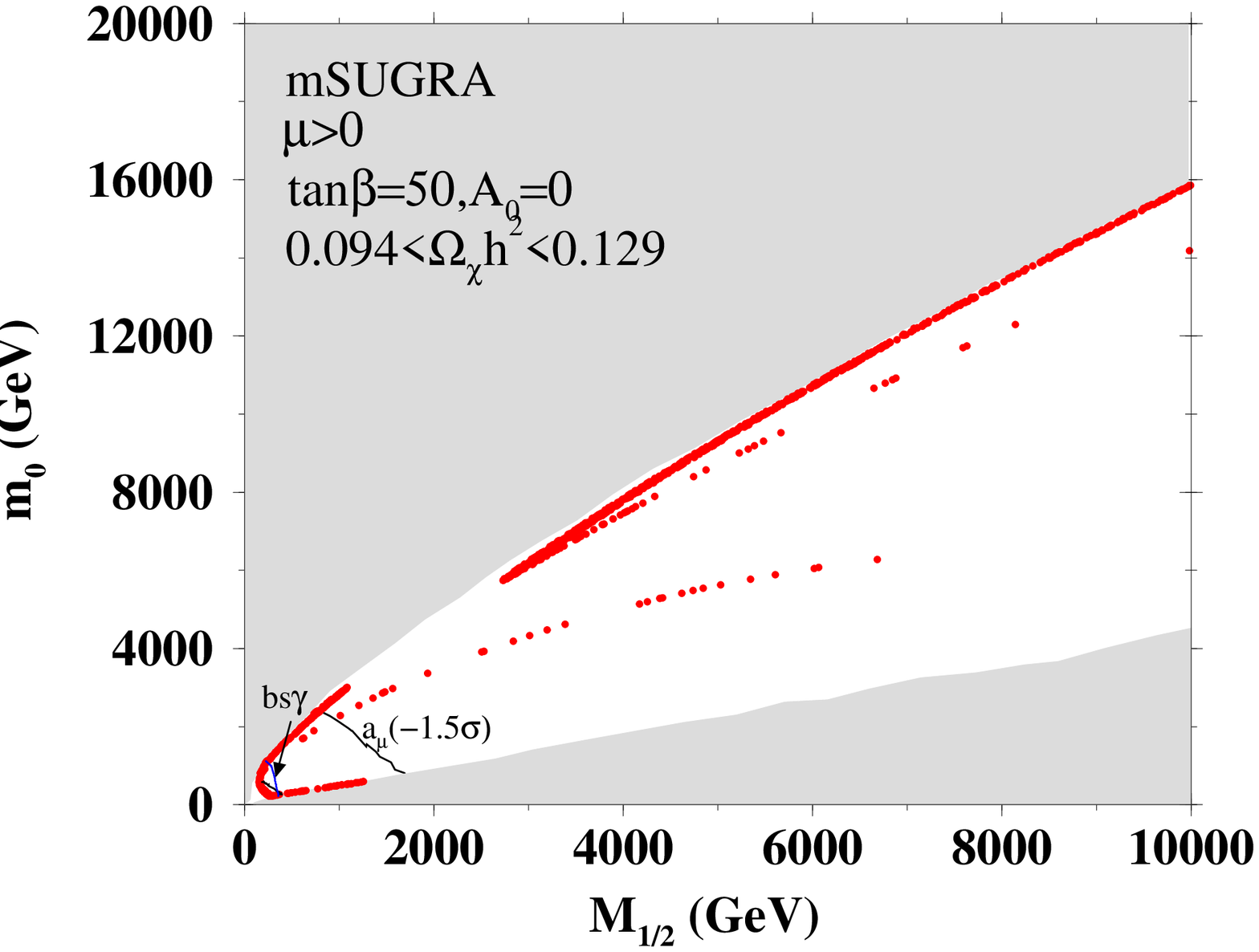} 
\end{minipage}}                       

%KEEP A LINE BLANCK AS ABOVE IN ORDER TO HAVE THE FOLLOWING TWO FIGS AT 
% THE BOTTOM 
\subfigure[Same as Fig.~\ref{t10_m0_mchi_msugra} 
except  $\tan\beta =50$.]{
\label{t50_m0_mchi_msugra}                      
\hspace*{-0.6in}                     
\begin{minipage}[b]{0.5\textwidth}                       
\centering
\includegraphics[width=\textwidth]{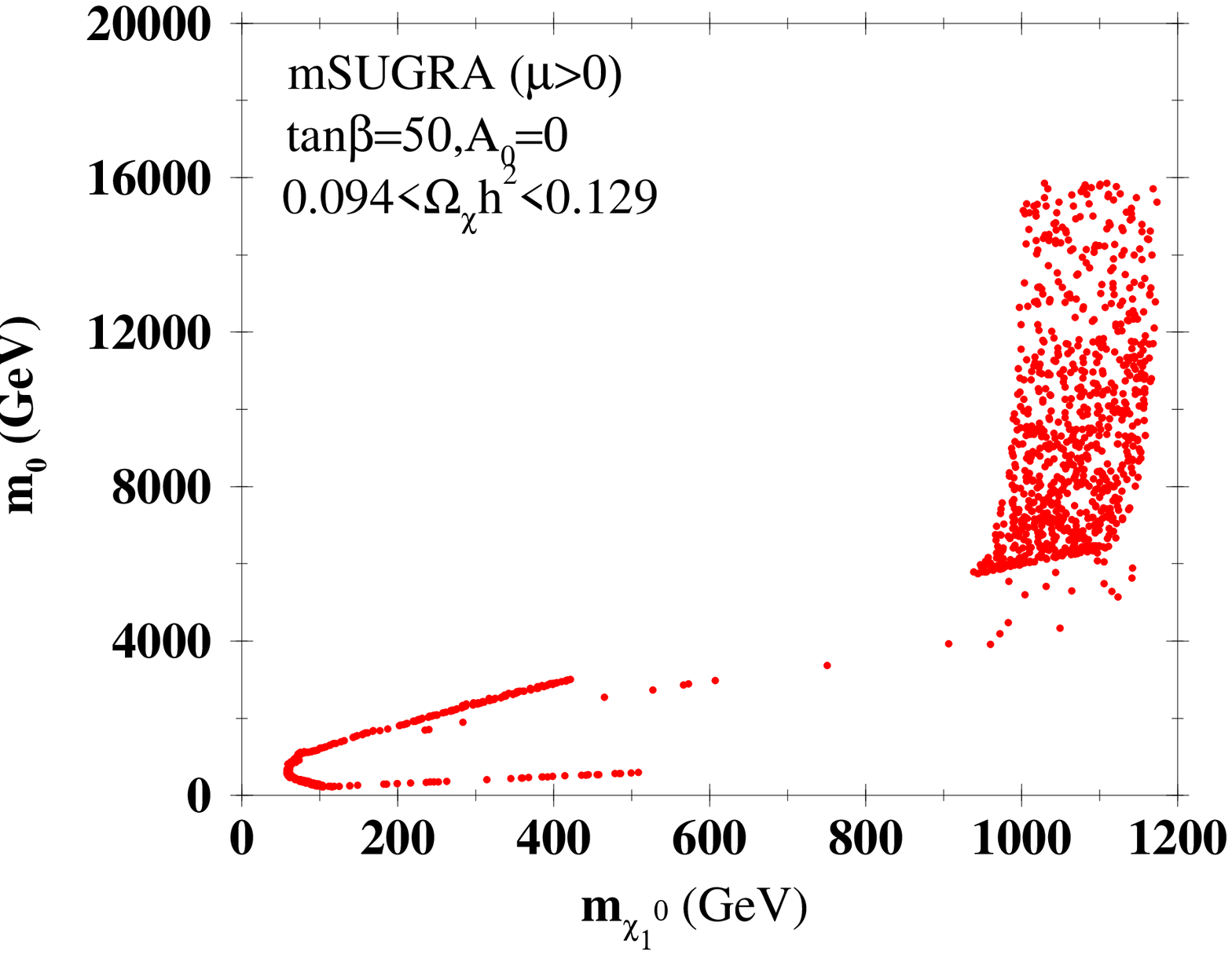}
\end{minipage}}
\hspace*{0.3in}                       
\subfigure[Same as Fig.~\ref{t10_mhalf_mchi_msugra} 
except $\tan\beta =50$.]{
\label{t50_mhalf_mchi_msugra}                       
\begin{minipage}[b]{0.5\textwidth}                       
\centering                      
\includegraphics[width=\textwidth]{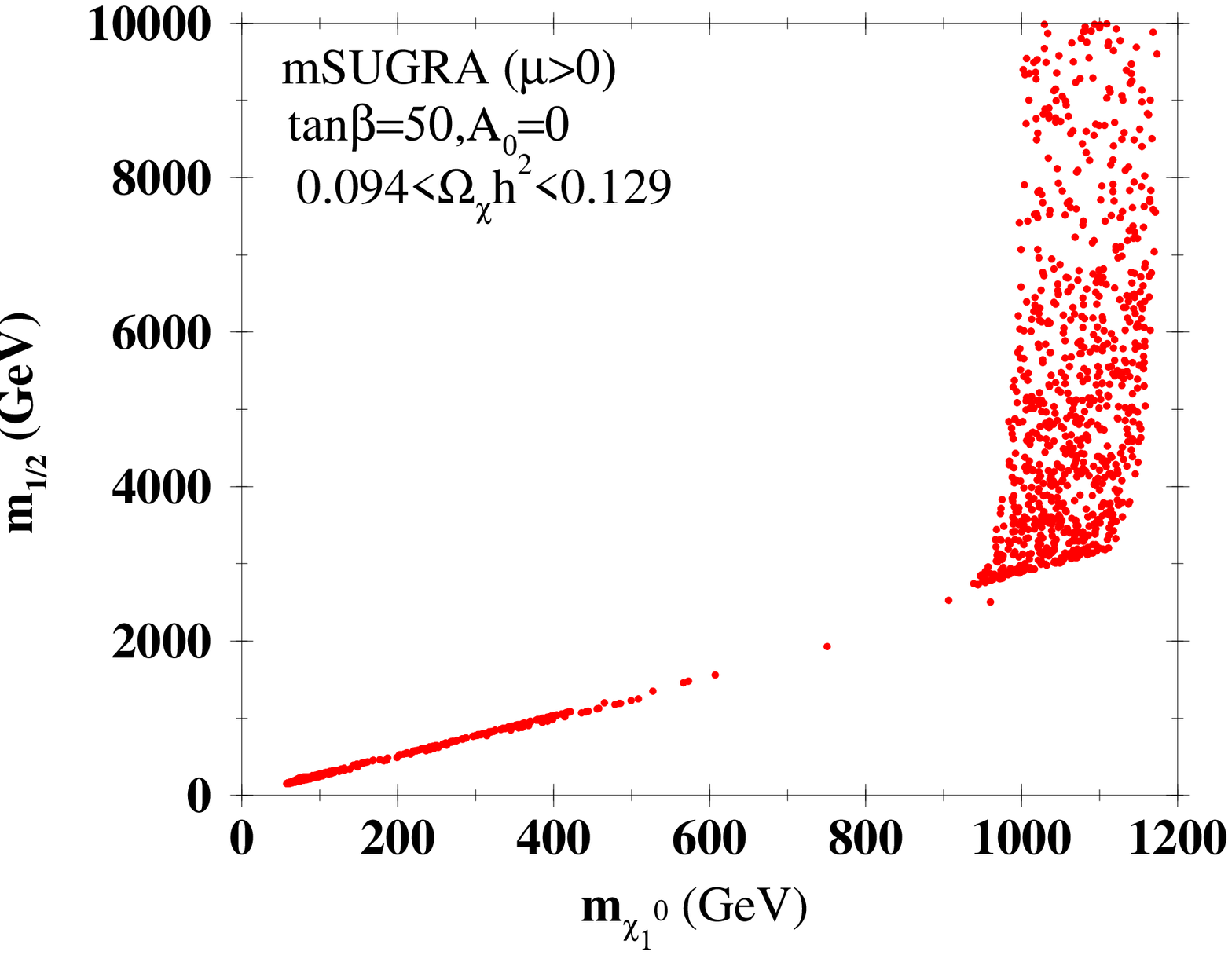}
\end{minipage}}                       
\caption{Relic density constraint and neutralino mass range for $\tan\beta=50$}
\label{tan50threefigs}
\end{figure}

In the analysis we include the 
$b\rightarrow s\gamma$ constraint\cite{cleo} and the $g_{\mu}-2$
constraint\cite{bnl2002}. The constraint arising from 
$B^0_{s}\rightarrow \mu^+\mu^-$ for large $\tan\beta$ is also 
discussed.  The analysis of 
Ref.\cite{narison} gives two estimates 
for the difference $a_{\mu}^{exp}-a_{\mu}^{SM}$:
These are [I] $a_{\mu}^{exp}-a_{\mu}^{SM}=1.7(14.2)\times 10^{-10}$\cite{davier,narison}
and [II]
$a_{\mu}^{exp}-a_{\mu}^{SM}=24.1(14.0)\times 10^{-10}$\cite{hagiwara,narison}.
These estimates 
also include corrections from scalar mesons to the muon anomaly 
computed in Ref.\cite{narison}.
Estimate [I] corresponds to essentially a perfect agreement and
does not put any effective upper limit constraints on the parameter
space.
In our analysis we consider a $1.5\sigma$ range around the central value of
 estimate [II], i.e., we choose  
$3.1\times 10^{-10}\leq (a_{\mu}^{exp}-a_{\mu}^{SM})\leq 45.1\times 10^{-10}$.
We attribute the difference to supersymmetry\cite{yuan}.
In Fig.~\ref{t10_m0_mhalf_msugra} we exhibit the allowed 
parameter space in the $m_0-m_{\frac{1}{2}}$ plane which satisfies
the relic density constraint consistent with Eq.(\ref{wmapeqn}) for the case
$\tan\beta =10$ and $\mu>0$. 
The filled dark circles indicate the regions which are consistent
with the relic density constraints. We note that this region includes
a lower branch  which is the conventional branch where the relic
density constraints are satisfied due to coannihilation with staus.
For the case of Fig.~\ref{t10_m0_mhalf_msugra} 
this extends to  $m_{\frac{1}{2}}$ of about 
800 GeV  and $m_{\chi_1^0}$ of about 300 GeV as can be seen more 
clearly from Fig.~\ref{t10_m0_mchi_msugra}.   
However, there is also an  upper branch where the
allowed values of  $m_{\frac{1}{2}}$ consistent with relic density
run up to the upper limit chosen i.e. 10 TeV. 
 The corresponding 
neutralino mass, however, runs up only to 1200 GeV because of the
phenomenon of inversion discussed in Sec.2. As can be seen from
Fig.~\ref{t10_m0_mchi_msugra} 
relic density constraints consistent with the WMAP constraints
can be satisfied in the inversion region for significantly large
values of the neutralino mass and values of $m_0$ up to 16 TeV.
The phenomenon of inversion can be seen more clearly in 
Fig.~\ref{t10_mhalf_mchi_msugra} where
points consistent with the WMAP constraints are exhibited in
the $m_{\frac{1}{2}}-m_{\chi_1^0}$ plane. The imposition of the $g_{\mu}-2$
constraint [II] eliminates all of the inversion region and much of 
the remaining region of the high zone of the hyperbolic branch.
 However, essentially all of the region 
allowed by the relic density constraints  is valid if we consider
the $g_{\mu}-2$ constraint [I].  
In Figs.~\ref{t30_m0_mhalf_msugra}, \ref{t30_m0_mchi_msugra},
\ref{t30_mhalf_mchi_msugra} 
we give an analysis similar to that of 
Figs.\ref{t10_m0_mhalf_msugra},
\ref{t10_m0_mchi_msugra}, and \ref{t10_mhalf_mchi_msugra}
except that  $\tan\beta =30$.
Similarly in
Figs.\ref{t50_m0_mhalf_msugra},
\ref{t50_m0_mchi_msugra}, and \ref{t50_mhalf_mchi_msugra}
we give an analysis similar to that of 
Figs.\ref{t10_m0_mhalf_msugra},
\ref{t10_m0_mchi_msugra}, and \ref{t10_mhalf_mchi_msugra}
except  $\tan\beta =50$.
For the cases of $\tan\beta =30$ and $\tan\beta =50$
the $b\rightarrow s\gamma$ 
constraint is also displayed. In these cases the region below the
curves labelled $b-> s\gamma$ is the disallowed region.

\begin{figure}
\hspace*{-0.6in}
\centering
\includegraphics[angle=0,width=\textwidth]{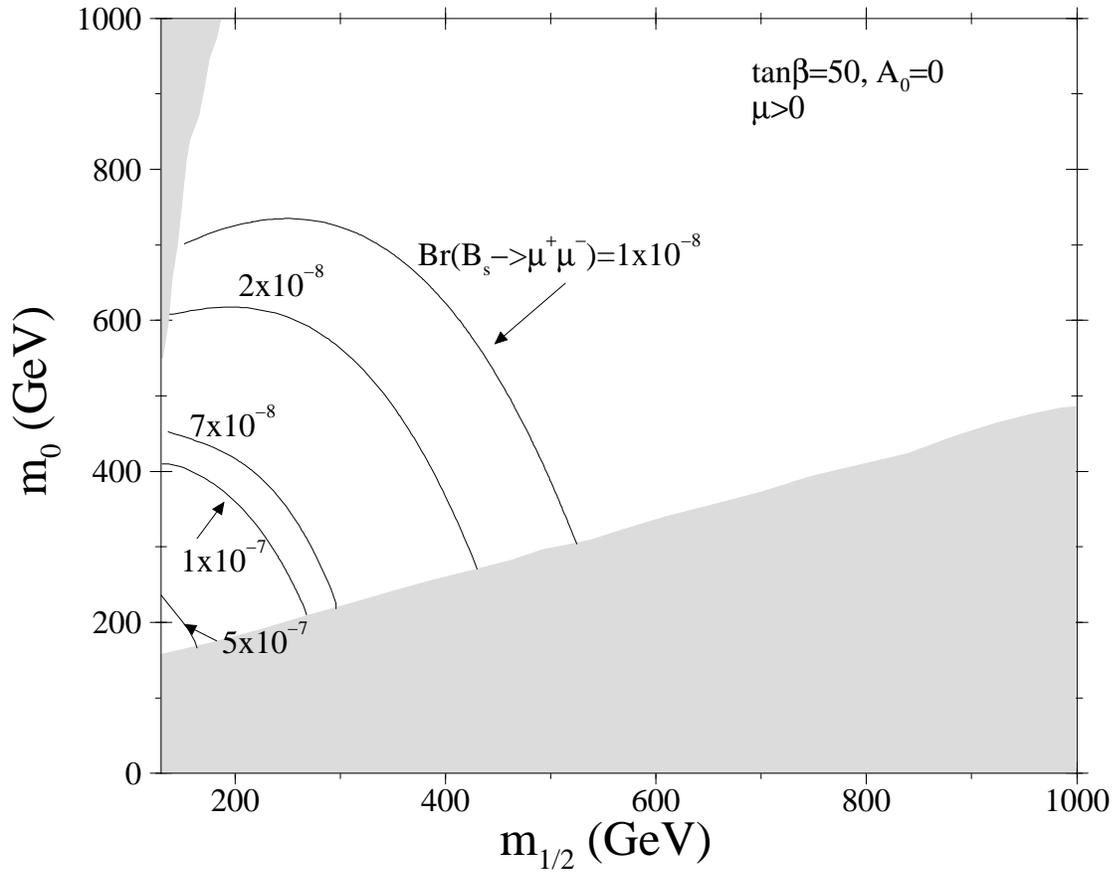}
\caption{\small A plot of the $B(B^0_s\rightarrow \mu^+\mu^-)$ constraint
in the $m_{\frac{1}{2}}-m_0$ plane.}
\label{bmumu}
\end{figure}

 For large $\tan\beta$ the constraint from 
$B^0_{s}\rightarrow \mu^+\mu^-$ is also of
interest\cite{gaur,dedes}. In the standard model the branching ratio
for this process is $B(\bar B^0_s\rightarrow \mu^+\mu^-)=(3.1\pm 1.4)
\times 10^{-9}$ ($V_{ts}=0.04\pm 0.002$) while the current limit 
from experiment is 
$B(\bar B^0_s\rightarrow \mu^+\mu^-)< 2.6 \times 10^{-6}$. 
The current estimates are that RUNII of the Tevatron will eventually 
increase the 
sensitivity for this process to the limit $10^{-8}$\cite{dedes}
which still falls short of reaching the branching ratio for this
process in the  standard model. However, it turns out that
in supersymmetry this branching ratio is dominated by the so called
counterterm diagram and the contribution from this diagram gives 
the branching ratio a dependence on $\tan\beta$ of $\tan^6\beta$ for 
large $\tan\beta$. As a consequence the $B(\bar B^0_s\rightarrow \mu^+\mu^-)$
branching ratio in 
supersymmetry can get larger than the standard model value by 
as much as a factor of $10^{3}$ which brings it within reach of
RUNII of the Tevatron. However, the $B(\bar B^0_s\rightarrow \mu^+\mu^-)$
branching ratio in supersymmetry is very sensitive to the 
sparticle spectrum and falls sharply as the
sparticle spectrum becomes heavy. In Fig.~\ref{bmumu} we give a plot
of the  $B(\bar B^0_s\rightarrow \mu^+\mu^-)$ constraint in the
$m_0-m_{\frac{1}{2}}$ plane.  We find that the current experimental
constraint on $B(\bar B^0_s\rightarrow \mu^+\mu^-)$ does not eliminate
any relevant part of the parameter space while 
$B(\bar B^0_s\rightarrow \mu^+\mu^-)=10^{-8}$ can explore the parameter 
space in $m_0$ up to 700 GeV and in $m_{\frac{1}{2}}$ up to about 
$500$ GeV. This mass range is far too small to have any influence
on the region of the hyperbolic branch we are focussing on in this
analysis. For this reason tbis constraint is not very effective 
in the present analysis.
 
\begin{figure}                       
\vspace*{-1.5in}
\subfigure[A plot of the neutralino-proton spin independent cross section
 $\sigma_{\chi_1^0 p}(SI)$ vs the neutralino mass for the allowed region
 of the parameter space for all the same input parameters and 
 constraints as in Fig.~\ref{t10_m0_mhalf_msugra}]{
\label{t10_sigmasi_msugra}                      
\hspace*{-0.6in}
\begin{minipage}[b]{\textwidth}                       
\centering
\includegraphics[width=0.65\textwidth]{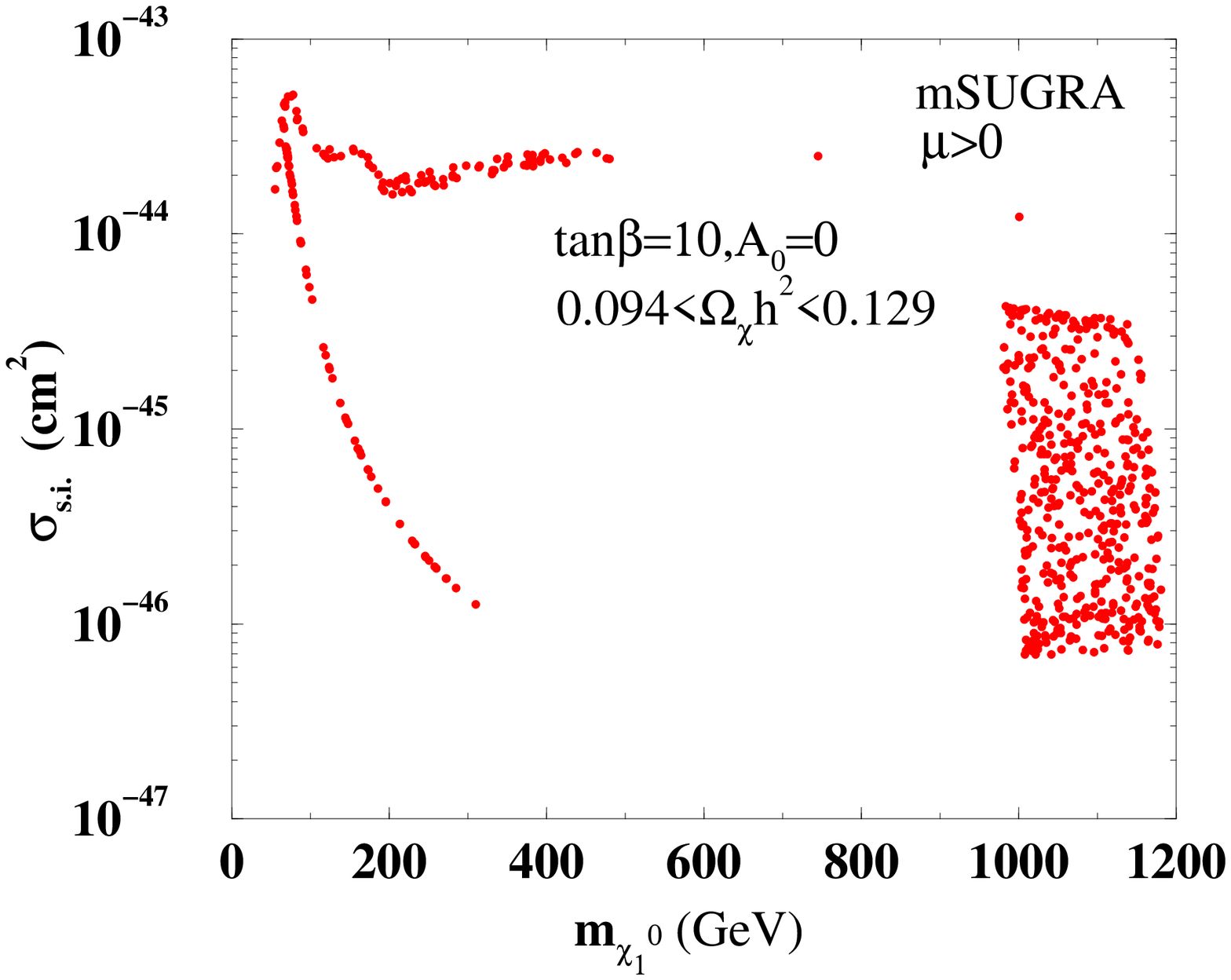}
\end{minipage}}                       

% 1 line gap given here above for the next fig to be at the bottom
\subfigure[A plot of the neutralino-proton spin dependent cross section
 $\sigma_{\chi_1^0 p}(SD)$ vs the neutralino mass for the allowed region
 of the parameter space for all the same parameters and 
 constraints as in Fig.~\ref{t10_m0_mhalf_msugra}]{
\label{t10_sigmasd_msugra}   
\hspace*{-0.6in}            
\begin{minipage}[b]{\textwidth}                       
\centering                      
\includegraphics[width=0.65\textwidth]{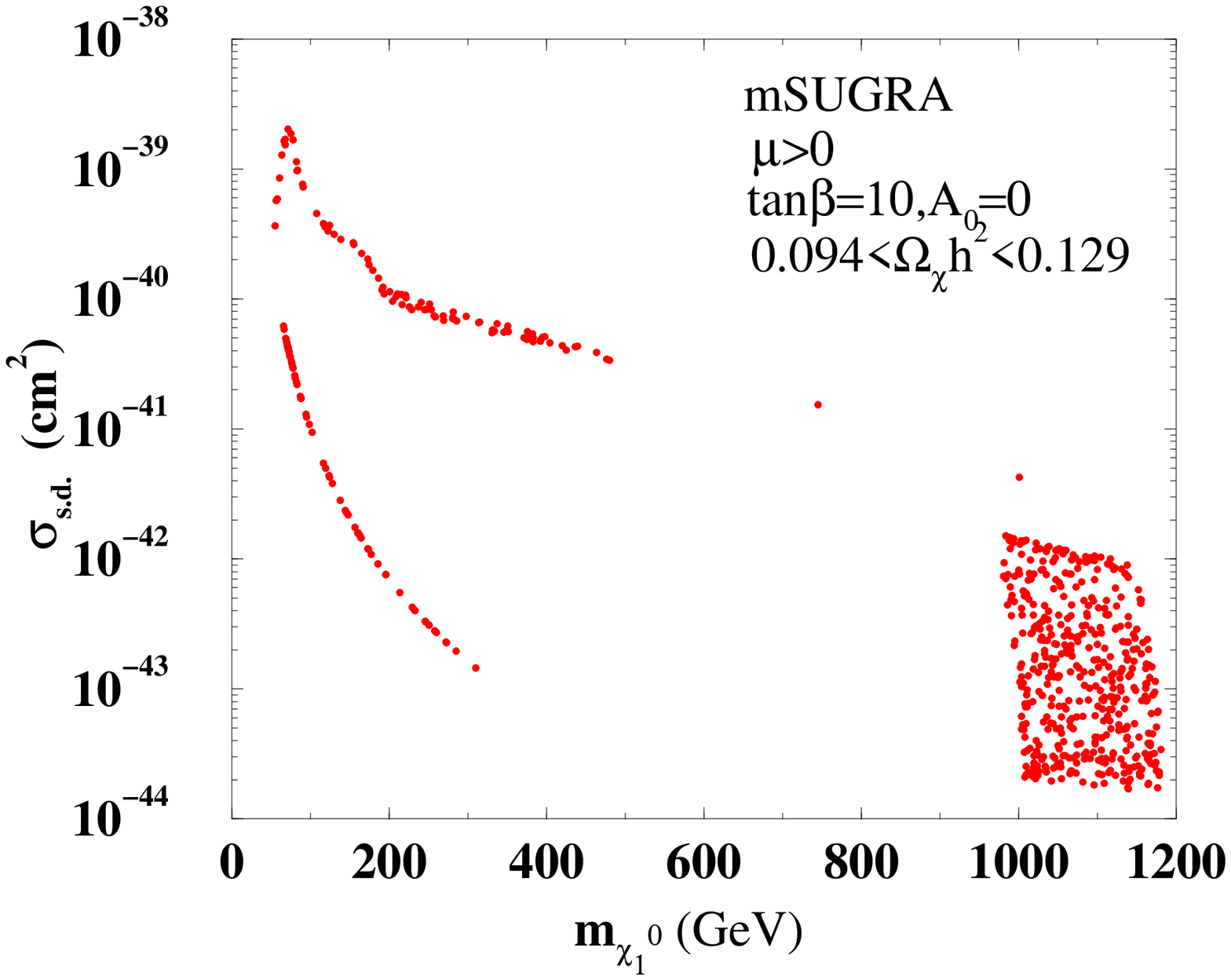}
\end{minipage}}                       
\caption{Spin Independent and Spin Dependent Cross Sections for 
$\tan\beta=10$}                      
\label{sigmafigstan10} 
\end{figure}

\begin{figure}                       
\vspace*{-1.5in}
\subfigure[Same as Fig.~\ref{t10_sigmasi_msugra} except 
$\tan\beta =30$.]{
\label{t30_sigmasi_msugra}                      
\hspace*{-0.6in}
\begin{minipage}[b]{\textwidth}                       
\centering
\includegraphics[width=0.65\textwidth]{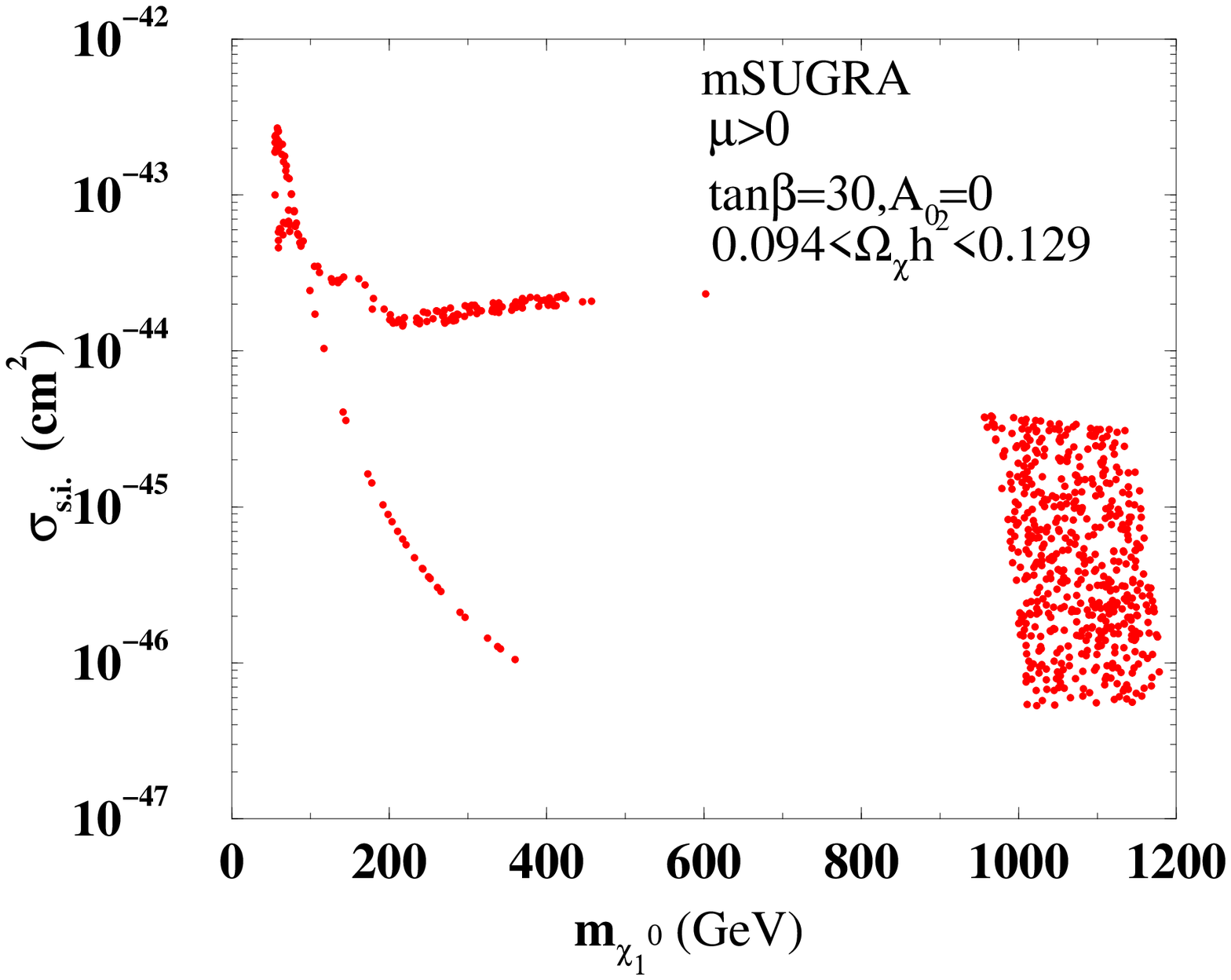}
\end{minipage}}                       

% 1 line gap given here above for the next fig to be at the bottom
\subfigure[Same as Fig.~\ref{t10_sigmasd_msugra}
 except $\tan\beta=30$.]{
\label{t30_sigmasd_msugra}   
\hspace*{-0.6in}            
\begin{minipage}[b]{\textwidth}                       
\centering                      
\includegraphics[width=0.65\textwidth]{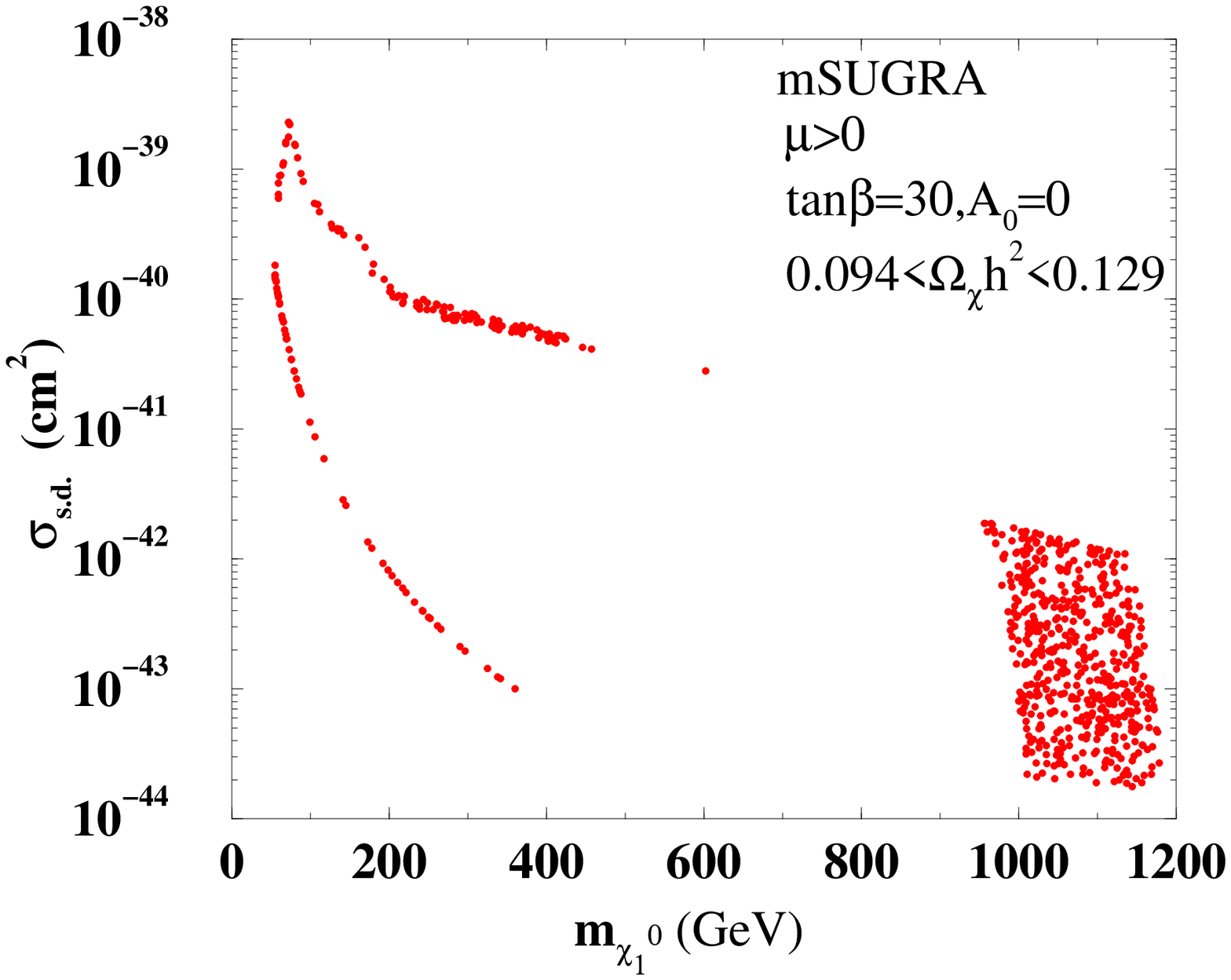}
\end{minipage}}                       
\caption{Spin Independent and Spin Dependent Cross Sections for 
$\tan\beta=30$}                      
\label{sigmafigstan30} 
\end{figure}

\begin{figure}                       
\vspace*{-1.5in}
\subfigure[Same as Fig.~\ref{t10_sigmasi_msugra} except 
$\tan\beta =50$.]{
\label{t50_sigmasi_msugra}                      
\hspace*{-0.6in}
\begin{minipage}[b]{\textwidth}                       
\centering
\includegraphics[width=0.65\textwidth]{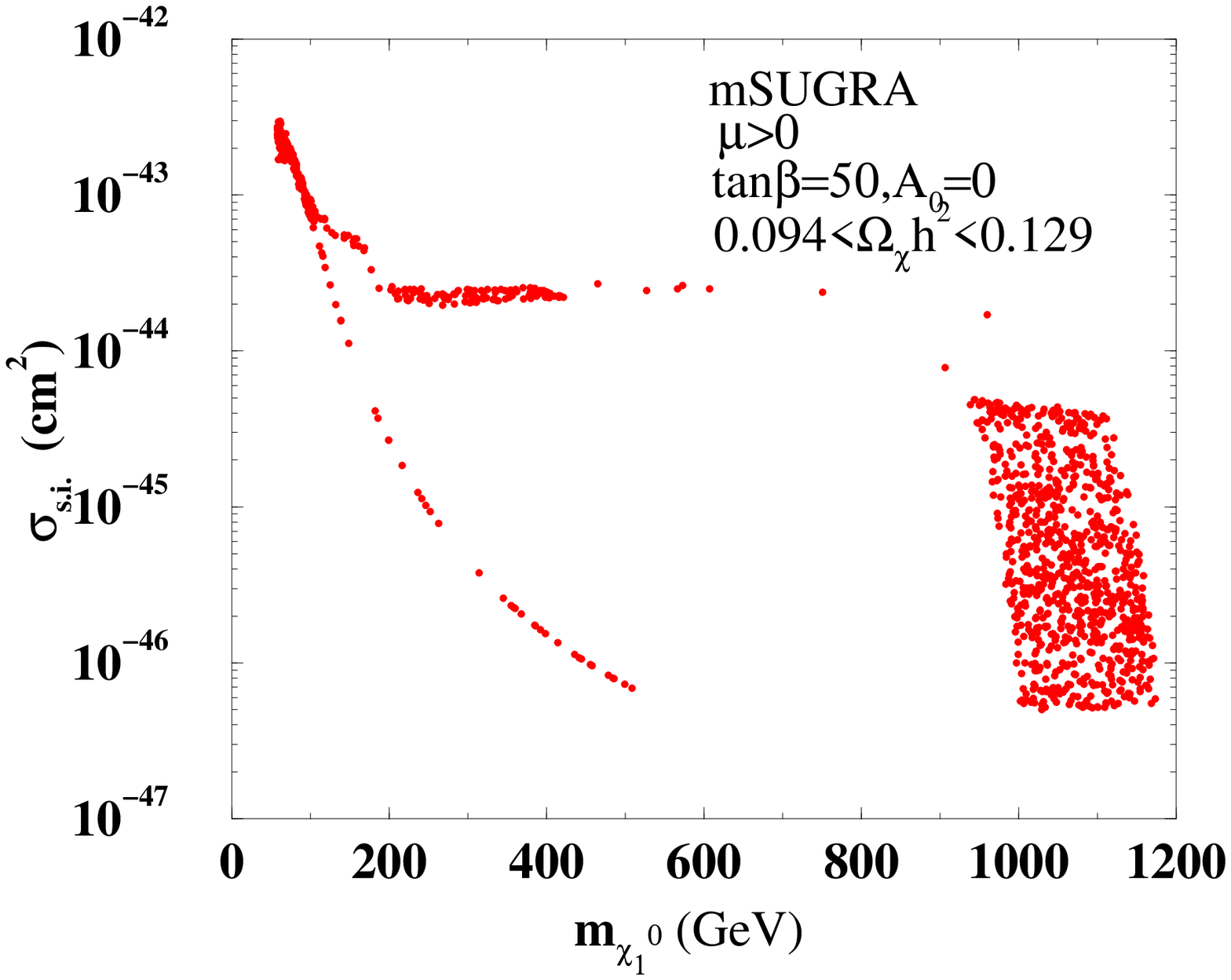}
\end{minipage}}                       

% 1 line gap given here above for the next fig to be at the bottom
\subfigure[Same as Fig.~\ref{t10_sigmasd_msugra}
 except $\tan\beta=50$.]{
\label{t50_sigmasd_msugra}   
\hspace*{-0.6in}            
\begin{minipage}[b]{\textwidth}                       
\centering                      
\includegraphics[width=0.65\textwidth]{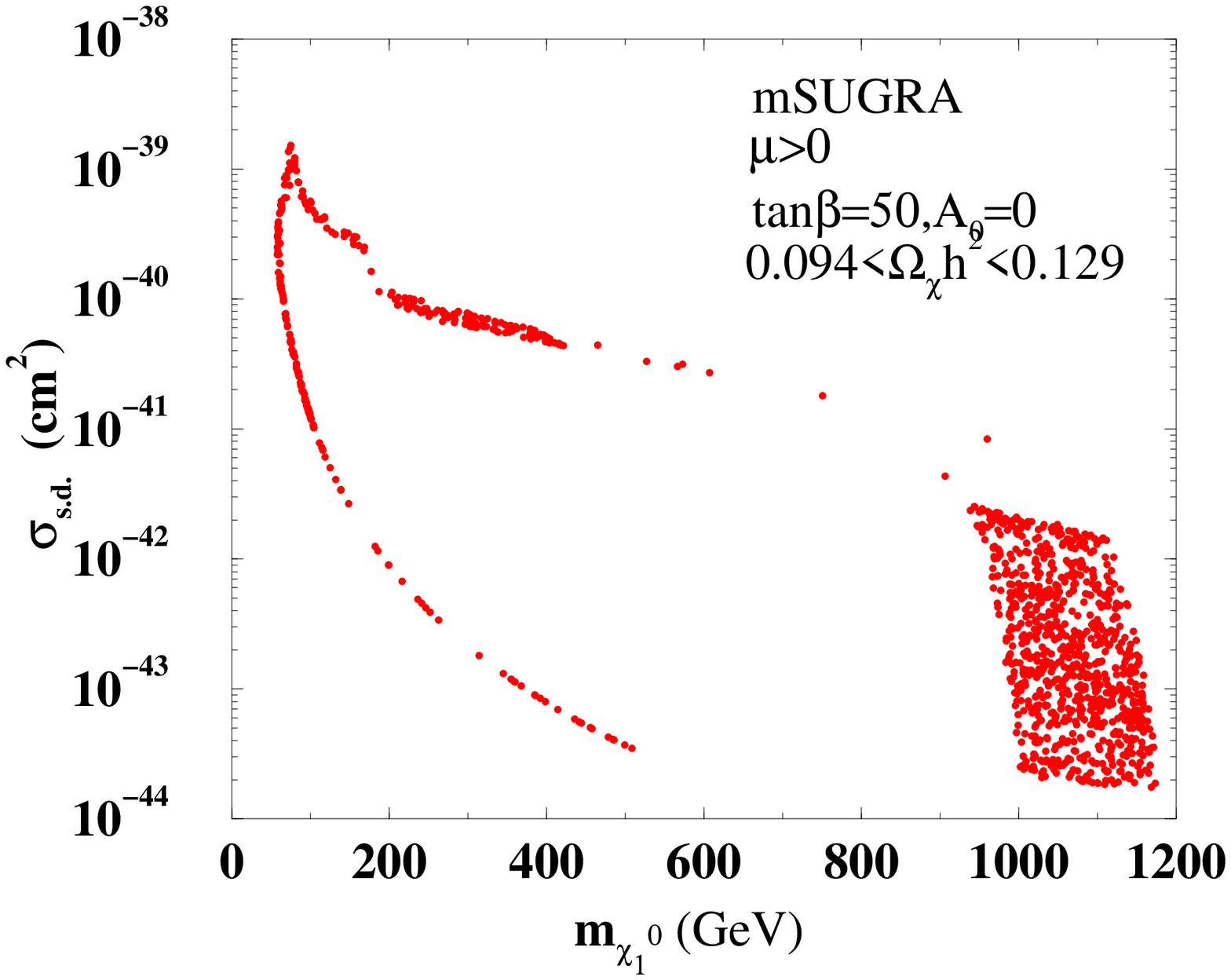}
\end{minipage}}                       
\caption{Spin Independent and Spin Dependent Cross Sections for 
$\tan\beta=50$}                      
\label{sigmafigstan50} 
\end{figure}

A quantity of great interest is the spin independent neutralino-proton 
cross section $\sigma_{\chi_1^0p}(SI)$ on which experimental
limits exist from the current dark matter experiments so that
$\sigma_{\chi_1^0p}(SI)\leq 10^{-42}cm^2$.   
In Fig.~\ref{t10_sigmasi_msugra} we  give a plot of
$\sigma_{\chi_1^0p}(SI)$  for $\tan\beta =10$ and $\mu>0$.
In Fig.~\ref{t10_sigmasi_msugra} the lower rapidly falling curve
that terminates at $m_{\chi_1^0}=300$ GeV is the branch on which
staus coannihilation occurs.  The upper curve arises from the low zone 
of the hyperbolic branch while the patch to the right is the one
that arises from the inversion region of the hyperbolic branch.
 For values of neutralino
masses below 300 GeV the $\sigma_{\chi_1^0p}(SI)$ cross section
arising from the upper curve in Fig.1(a)  is much larger than the one 
arising from the lower branch where the relic density
constraints are satisfied due to neutralino-stau coannihilation. 
We also note that in Fig.~\ref{t10_sigmasi_msugra} the patch to the
right indicates that the scalar cross sections are quite significant
even though one is in the inversion region. 
Thus although the direct detection of supersymmetry in the inversion
region is  more difficult, the  neutralino-proton scalar crosss  are
still substantial. 
In the  future dark matter detectors\cite{genius} will be able to achieve
a sensitivity of up to $10^{-45}cm^2$. We note that a significant
part of the parameter space of Fig.~\ref{t10_sigmasi_msugra}
will be probed by these detectors. 
In Fig.~\ref{t10_sigmasd_msugra} we give a plot of
the  spin dependent neutralino-proton cross section 
$\sigma_{\chi_1^0p} (SD)$  for $\tan\beta =10$ $\mu>0$.
A comparison of Fig.~\ref{t10_sigmasi_msugra}
and Fig.~\ref{t10_sigmasd_msugra} shows that the spin dependent
cross section is typically much larger than the spin dependent 
cross section by 3-4 orders of magnitude.
 A similar analysis for the 
case $\tan\beta =30$ is given in 
Fig.~\ref{t30_sigmasi_msugra}
and Fig.~\ref{t30_sigmasd_msugra}
while for the case $\tan\beta =50$ is given in 
Fig.~\ref{t50_sigmasi_msugra}
and Fig.~\ref{t50_sigmasd_msugra}.
 The conclusions for these cases are very similar to the
  conclusions drawn from  Fig.~\ref{t10_sigmasi_msugra}
and Fig.~\ref{t10_sigmasd_msugra}.  Based on these analyses
one finds that for $\tan\beta\leq 50$, 
the neutralino mass range consistent with the
WMAP constraints on the branch corresponding to neutralino-stau 
coannihilation is $m_{\chi_1^0}\leq 500$ GeV
 and  $m_{\chi_1^0}\leq 1200$ GeV 
 for the high zone of the hyperbolic branch where the relic density
 constraints are satisfied due to coannihilation with the next
 to lightest neutralino and the light chargino.
These constraints remain intact under the imposition of the 
$g_{\mu}-2$ constraint [I] but the constraint arising from 
the inversion region of the hyperbolic branch 
is removed by imposition of the $g_{\mu}-2$ constraint [II].

\section{WMAP Constraints and Discovering SUSY at accelerators}
The analysis of Sec.3 shows that Eq.(\ref{wmapeqn}) 
constraints the parameter
space very strigently. 
For the usually explored parameter space of minimal supergravity where 
relic density is satisfied in the region of neutralino-stau 
coannihilation as well as in the low zone of the hyperbolic branch 
with a moderate 
amount of higgsino in the lsp (i.e. without inversion), 
one finds that the neutralino mass now
 has an upper limit of about 500 GeV for $\tan\beta\leq  50$ 
 and $m_0$ lies in the few hundred GeV range.  
For  this  
case the corresponding sparticle spectrum should all be  accessbile
at the LHC and perhaps some of it may be accessible at RUNII at the
Tevatron. Also there are some interesting signals for this  branch
at the NLC\cite{kamon}. 
However, on the inversion region of the 
hyperbolic branch of radiative breaking of the
electroweak symmetry, $m_0$ and $m_{\frac{1}{2}}$ can get as large as
10 TeV or even higher. In this case the squarks and the sleptons 
would lie in the several TeV region and hence they would be 
beyond the reach of even the LHC.  The light particles in this
scenario will be the two  lightest neutralinos
and the light chargino. However, the signals
for their detection would be significantly different than for 
the normal scenarios. Specifically, in the inversion region of the
hyperbolic branch  the mass differences among
$\chi_1^0$, $\chi_1^{\pm}$ and $\chi_2^0$ are so small that the 
usual signals discussed for the detection of supersymmetry would not
apply\cite{feng}. 

Situations of the type above have been discussed before in 
Ref.\cite{cdg} in the context of string models and in 
Ref.\cite{fmrss} in the context of Wino lsp scenarios
while the experimental search for charginos mass-degenerate with the
lightest neutralinos has been analyzed in Ref.\cite{delphi}. 
Here the mass scales are significantly different. Thus, for example,
in the analyzes of Ref\cite{fmrss} the mass difference of the
chargino and the nearly degenerate neutralino is in the range of 
$O(100)$ MeV which allows for charged particle tracks in the
detector of the order of few centimeters arising from the decay of
the chargino to neutralino such as 
$\chi_1^{+}\rightarrow \chi_1^0 l^+\nu_l$, and 
$\chi_1^{+}\rightarrow \chi_1^0 l^+\pi^+$.
In the present scenario the  
chargino and neutralino masses are in the several hundred GeV to
1-2 TeV range and their mass difference lie in the range of 
1-10 GeV. The mass differences are such that the chargino will 
always decay in the detector and the track length will be 
too small to be visible. 
Further, the conventional trileptonic signal\cite{trilep} would
yield leptons with energies only in the few GeV region to 
provide a useful signal at the LHC\cite{cms}. In Ref.\cite{cdg} it 
is argued that charginos nearly degenerate with neutralinos may 
be observable in $e^+e^-$ 
colliders via observation of hard photons in the process
$e^+e^-\rightarrow \gamma\chi_1^+\chi_1^-$. However, 
a more detailed analysis for the detection of supersymmetry
in collider experiments is needed for the scenario discussed
here. On the whole, the prospects for the detection of SUSY 
signals at accelerators in this scenario look difficult. 
On the other hand
quite interestingly this scenario does provide a sufficient
amount of dark matter to populate the universe and a part of the
parameter space of this branch does yield spin independent
neutralino-proton cross sections which lie in the range of 
observability of dark matter detectors. We emphasize that much
of the high zone of the hyperbolic branch and specifically all of the inversion 
region on the hyperbolic branch can be eliminated  if the 
$g_{\mu}-2$ constraint [II] holds. However, the high zone of the
hyperbolic branch would not be significantly constrained if the $g_{\mu}-2$ 
constraint [I] holds. This points to the importance of getting
an unambiguous determination of the 
leading order (LO) hadronic correction to  $g_{\mu}-2$.

We comment now briefly on the relation of the hyperbolic branch to 
the focus point region\cite{fmm}. As discussed in Sec.1 we showed that
one can find solutions to radiative breaking of the electroweak symmetry 
where $m_0$ and $m_{1/2}$ can get large while $\mu$ remains fixed and relatively 
small. These solutions constitute the hyperbolic branch. 
A part of this region also includes the so called focus point region.
Thus the focus point region is limited to relatively small values of 
$m_{1/2}$ and consequently $m_0$ is also limited from getting very
large because of the radiative symmetry breaking constraint relative
to the case of the hyperbolic branch. Thus the focus point region (FP)
is truly a subset of the hyperbolic branch (HB). A further discussion
of this point can be found in Ref.\cite{hb/fp} where the 
acronym HB/FP is used to describe this region. 
\section{Conclusion}
The recent WMAP determination of the cosmological parameters, 
specifically $\Omega_m h^2$ and $\Omega_b h^2$, to a much better
accuracy than earlier determinations has important consequences
for the observation of supersymmetric dark matter and also for the
direct detection of supersymmetry. In our analysis we have 
identified the difference $\Omega_m h^2- \Omega_b h^2$ as 
arising from relic neutralinos and analyzed this possibility
within mSUGRA. One finds that for the region of the parameter space
where the relic density constraints are satisfied due to
the neutralino-stau coannihilation,  
the neutralino mass limit is now reduced to 
$m_{\chi_1^0}\leq 500$ GeV for $\tan\beta \leq 50$. 
The spectrum in this case will all be accessible at the LHC with
the possibility of some sparticles also being accessible at 
RUNII of the Tevatron. Also some interesting signals may arise in 
this case at the NLC.
On the high zone of the hyperbolic branch including the inversion
region,
 the WMAP constraints are satisfied remarkably to a very
high value of the neutralino mass, i.e., up to 
$m_{\chi_1^0}\leq 1200$ GeV for $\tan\beta \leq 50$.  
The satisfaction of the relic density even 
 for such large neutralino masses comes about because of 
 coannihilation processes exhibited in Eq.(\ref{charginocoani}). 
 As discussed in Sec.2, in the high zone of   
the hyperbolic branch $m_0$ and $m_{\chi_1^0}$ can get very
large and some of the sparticle spectrum may lie outside the reach
of even the LHC. Thus the squarks, sleptons and gluinos may
be too massive to be accessible even at the LHC. Thus the direct 
observation of SUSY would be very challenging if the inversion 
region of the hyperbolic branch is realized. 
In this region the only light  
   particles, aside from the light Higgs boson 
 $h^0$, are the sparticles $\chi_1^{\pm}, \chi_1^0, \chi_2^0$. 
 The mass splittings among them are typically O(10) GeV and thus
 their detection poses a challenge.
Luckily much of the hyperbolic branch and all of the inversion region
of the hyperbolic branch can be eliminated by a $g_{\mu}-2$ 
signal. This is what happens when we impose the $g_{\mu}-2$ 
constraint [II]. However, imposition of the $g_{\mu}-2$ constraint [I]
essentially leaves all of the region of the hyperbolic branch 
including the inversion region intact. This points to the need to
achieve an unambiguous $g_{\mu}-2$ constraint by reducing the
errors in the leading order (LO) hadronic contributions. 
We also computed the spin independent 
 neutralino proton  cross section $\sigma_{\chi_1^0 p} (SI)$
 and found that it lies in the range $10^{-46}-10^{-43} cm^2$. A significant 
 part of this range will be accessible
 to the future dark matter experiments\cite{genius,cline}. 
 Implications of WMAP constraints for supersymmetry have  
 also been reported in  Ref.\cite{elliswmap}. \\
  \noindent
 {\bf Acknowledgments}\\ 
 Conversations with H. Baer, J. Feng and D. Wood related to topics discussed
 here are acknowledged.
  This research was supported in part by NSF grant  PHY-0139967

%%%%% FIGURES %%%%%%%%%%%%%%%%%%%%%%%%%%%%%%%%%%%
\newpage

%%%%%%%%%%%%%%%%%%%%%%%%%%%%%%%%%%%%%%%%%%%%%%%%%%%%%%%%%%%%%%%%%%%%%%%%%%
\newpage

%%%%%%%%%%%%%%%%%%%%%%%%%%%%%%%%%%%%%%%%%%%%%%%%%%%%%%%%%%%%%%%%%%%%%%%%%%
%%%%%%%%%%%%%%%%%%%%%%%%%%%%%%%%%%%%%%%%%%%%%%%%%%%%%%%%%%%%%%%%%%%%%%%%%%
\newpage

%%%%%

%%%%%%%%%%%%%%%%%%%%%%%%%%%%%%%%%%%%%%%%%%%%%%%%%%%%%%%%%%%%%%%%%%%%%%%%%%
\newpage

%%%%%%%%%%%%%%%%%%%%%%%%%%%%%%%%%%%%%%%%%%%%%%%%%%%%%%%%%%%%%%%%%%%%%%%%%%
%%%%%%%%%%%%%%%%%%%%%%%%%%%%%%%%%%%%%%%%%%%%%%%%%%%%%%%%%%%%%%%%%%%%%%%%%%
\newpage

%%%%%%%%%%%%%%%%%%%%%%%%%%%%%%%%%%%%%%%%%%%%%%%%%%%%%%%%%%%%%%%%%%%%%%%%%%
\newpage

\end{document}